\documentclass{llncs}
\usepackage[utf8]{inputenc}
\usepackage{graphicx}
\usepackage{xspace}
\usepackage{url}
\usepackage{listings}
\usepackage{fontawesome}

\newcommand{\eg}{\emph{e.g.,}\xspace}
\newcommand{\ie}{\emph{i.e.,}\xspace}
\newcommand{\etal}{\emph{et al.,}\xspace}
\newcommand{\ct}[1]{{\textsf{#1}}\xspace}

\newcommand{\famixng}{\textsc{FamixNG}\xspace}
\newcommand{\modmoose}{\textsc{ModMoose}\xspace}

\begin{document}

\title{Modular Moose:  A new generation \\ software reverse engineering environment}
\author{Nicolas Anquetil\inst{1}\textsuperscript{\faEnvelopeO}\orcidID{0000-0003-1486-8399} \and Anne Etien\inst{1}\orcidID{0000-0003-3034-873X}
 \and Mahugnon H. Houekpetodji \inst{1,3} \and Benoit Verhaeghe\inst{1,4}
 \and Stéphane Ducasse\inst{2}\orcidID{0000-0001-6070-6599} \and Clotilde Toullec\inst{2}
 \and Fatiha Djareddir \inst{3} \and Jerôme Sudich \inst{3}
 \and Moustapha Derras\inst{4}
}

\institute{
Universit\'{e} de Lille, CNRS, Inria, Centrale Lille, UMR 9189 -- CRIStAL, France\\
\email{nicolas.anquetil@univ-lille.fr}\\
 \and
Universit\'{e} de Lille, Inria, CNRS,  Centrale Lille, UMR 9189 -- CRIStAL, France\\
\and 
CIM, Lille, France\\
\and
Berger-Levrault, Montpellier, France
}

\maketitle

\begin{abstract}
Advanced reverse engineering tools are required to cope with the complexity of software systems and the specific requirements of numerous different tasks (re-architecturing, migration, evolution).
Consequently,  reverse engineering tools should adapt to a wide range of situations. Yet, because they require a large infrastructure investment,  being able to reuse these tools is key.
Moose is a reverse engineering environment answering these requirements.
While Moose started as a research project 20 years ago, it is also used in industrial projects, exposing itself to all these difficulties.
In this paper we present \modmoose, the new version of Moose.
\modmoose revolves around a new meta-model, modular and extensible; a new toolset of generic tools (query module, visualization engine, ...); and an open architecture supporting the synchronization and interaction of tools per task.
With \modmoose, tool developers can develop specific meta-models by reusing existing elementary concepts, and dedicated reverse engineering tools that can interact with the existing ones. 
\end{abstract}

\section{Introduction}

As software technologies evolve, old systems experience a growing need to follow this evolution.
From the end-user point of view, they need to offer functionalities entirely unforeseen when they were first conceived.
From the developer point of view, they need to adapt to the new technologies that would allow one to implement these functionalities \cite{Deme02a}.

Given the size of these systems and lack of resources in the industry, such evolution can only happen with the help of automated tooling \cite{Bell98a,Kien10a}.
Concurrently, because such tooling requires a large infrastructure investment, it must be generic and reusable across technologies and reverse engineering tasks.
This tooling needs to cope with the following problems:
\begin{itemize}
\item \emph{Diversity of languages and analyses.}
Many programming languages and versions of such languages are used in the industry.
Meta-modeling was proposed to cope with that diversity, but it does not solve all problems.
Software reverse engineering requires to represent source code with a high degree of details that are specific to each programming language and to the reverse engineering tasks themselves.
How to model the different needs for details while remaining generic is an issue meta-models have not tackled yet.

\item \emph{Sheer amount of data.}
Tools are particularly useful for large systems (millions of lines of code).
The size of the analyzed systems stresses the modeling capabilities of the tools and the efficiency of the analyses.

\item \emph{Reverse engineering task diversity.}
The evolution needs are numerous, from restructuring a system towards a micro component architecture \cite{Brag20a}, to migrating its graphical user interface \cite{Verh19aShort}, evolving its database \cite{Delp20aShort}, or cleaning the code \cite{Anq19aShort}.
This calls for various tools (query module, visualizations, metrics, navigation, analyses) that must integrate together to answer each need still acting as a coherent tool.

\item \emph{Specific tasks require specific tools.}
Some tools can be useful in various situations, but specialized tools are also needed\footnote{According to Bruneliere \cite{Brun14c}, the plurality of reengineering projects requires adaptable/portable reverse engineering tools.}.
If a tool is too specific to a technology, its advantages are lost when working with others.
It is important that they can be easily added and integrated into a reverse engineering environment. 

\end{itemize}

We report here our experience with Moose, a software analysis platform  answering to the basic needs of reverse engineering, reusable in different situations, extensible and adaptable to new needs \cite{Nier05c}.
Moose was initiated as a research project and still fulfills this role. But it is also used in a number of industrial projects \cite{Anq19aShort,Delp20aShort,Verh19aShort}.

Our redesign of Moose has the following goals:
(1) the ability to develop specific meta-models by reusing existing elementary concepts,
(2) the ability to develop dedicated reverse engineering tools by configuring and composing existing tools or developing new ones,
(3) the ability to seamlessly integrate new tools in the existing reverse engineering environment.
\modmoose, the new version of Moose described in this article, revolves around a modular extensible meta-model, a new toolset of generic tools (\eg query module, visualization engine) and an open architecture supporting the synchronization and interaction of tools.

The contributions of this article are
(1) the description of \modmoose new generation reverse engineering environment,
(2) \famixng supporting the definition of new meta-models based on the composition of elementary concepts implemented using stateful traits, and
(3) a bus architecture to support the communication between independent and reusable tools. 

In the following sections, we discuss the difficulties of having a generic, multi-language, software meta-model (Section~\ref{sec:metamodel}) and how we solved them with \famixng (Section~\ref{sec:famixNG}). Then, we briefly present the new generic tools and how the new open architecture of \modmoose supports the collaboration of group of tools and their synchronization (Section~\ref{sec:moose-ide}).
We conclude in Section~\ref{sec:conclu}.

\section{Software Representation and Reverse Engineering Tool Challenges}
\label{sec:metamodel}

To analyze code, tools need to represent it. Such a representation should be \emph{generic} (programming languages are all build around similar concepts that should be reusable), support \emph{multi-language}, and \emph{detailed} (a proper analysis must pay attention to little, meaningful, details).

Each one of these challenges may be tackled by meta-models.
For example, existing IDEs (Integrated Development Environments) like Eclipse, or software quality tools (\eg CAST, SonarQube) offer tools based on an internal representation of the source code.
But in Eclipse, the model is not the same for version management and source code highlighting.
In that way, they can represent with enough details the analyzed systems.
The Lightweight Multilingual Software Analysis approach proposed by Lyons \etal uses different parsers (one for each programming language analyzed) to populate their \emph{different} models: ``variable assignment locations'', or ``reverse control flow''. 
The challenge is still how to support the development of multiple dedicated and specific meta-models for different languages in a tractable way.

\subsection{Problem: Handling the Diversity of Languages}

Many publications highlight the need, for modern software analysis tools, to deal with multi-language systems (\eg \cite{Deme99d,Egyed00a,Leth04a,Lyon18a,Maye17a}).
The generality and reusability of the analysis tools depend on their ability to work with all these languages.

Meta-models were supposed to solve this issue. Naively, one could hope that a single ``object-oriented meta-model'' would be able to represent programs in all OO programming languages and another ``procedural meta-model'' would represent all procedural programming languages.

Thus, for example, in the Dagstuhl Middle Meta-model \cite{Leth04a} ``the notions of routine, function and subroutine are all treated the same.''
But in practice, one soon realizes that ``various programming languages have minor semantic differences regarding how they implement these concepts'' (\cite{Leth04a} again.)
To be meaningful and allow precise analyses, the meta-models must represent these little differences and the tools must ``understand'' their specific semantics.
As a consequence, Washizaki \etal \cite{Wash16a} report that ``even if a meta-model is stated to be language independent, we often found that it actually only supports a very limited number of languages.''
A good meta-model must represent all details to allow meaningful analyses.
Software is not linear, small details may have huge impacts.
Reverse engineering is about abstracting from the source code, but also sticking tightly to it because one does not know in advance what ``details'' are making a difference for different tasks.

Visibility rules are a concrete example of the complexity to represent varying details for apparently universal concepts: Abstracting the API of a class requires understanding the visibility rules of classes, methods, and attributes.
They are not exactly the same for Java (\ct{public}, \ct{protected}, \ct{private}, \ct{default package}, and now \ct{export}) or C++ (\ct{public}, \ct{protected}, \ct{private}, and \ct{friend}).
And the tools need to be aware of each different semantics to make the proper inferences.

Other problems are raised by the containment relationship (\ie ownership).
Many meta-models assume a hierarchical containment tree: all entities (except the root entity) are owned by exactly one parent entity.
However, in C\#, class definitions may be scattered over several files (\emph{partial} definitions), and in Objective-C or Smalltalk a package may add methods to a class defined in another package (extension methods).
In these contexts, the notion of single owner is less consensual.

Our experience designing and using a programming language meta-model, Famix, for more than 15 years showed that we need dedicated meta-models for each language.

\subsection{Single inheritance: Tyranny of the dominant decomposition}

Moose was based on a single meta-model (Famix \cite{Deme99dShort,Deme01y,Nier05c}) extensible with plugins.
It did allow to model many different programming languages\footnote{Some languages are only partially supported} (Ada, C, C++, FORTRAN, 4D\footnote{\url{http://en.wikipedia.org/wiki/4th_Dimension_(software)}}, Java, JavaScript, Mantis\footnote{\url{http://www.lookupmainframesoftware.com/soft_detail/dispsoft/339}}, Pharo, PHP, PostgresSQL/PgPlSQL, PowerBuilder\footnote{\url{http://en.wikipedia.org/wiki/PowerBuilder}}), and even other sources of information (\eg SVN logs).
But extending Famix was an awkward process since it relied on a single inheritance tree of concepts.

To address language variety, the choice was to have core entities representing ``typical features'' of programming languages (\eg function, methods, package, class) and extend this core for language specific entities (\eg the macros in C).
The Dagstuhl Middle Meta-model \cite{Leth04a} made the same choice.

\begin{figure}
\begin{center}
\includegraphics[width=0.33\linewidth]{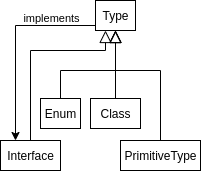}
\hfill
\rotatebox{90}{\makebox[4cm]{\Huge\dotfill}}
\hfill
\includegraphics[width=0.50\linewidth]{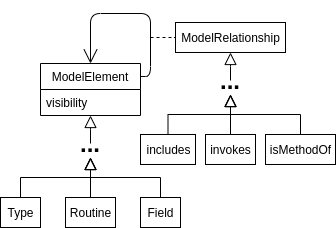}
\caption{Examples of the Dominant Decomposition problem in the Dagstuhl Middle Meta-model (left) and in Famix (right)}
\label{fig:tyranny}
\end{center}
\end{figure}

However this was only possible through a careful definition of the meta-model and the hierarchy of modeled entities.
In practice, Famix core was tailored towards Java and Smalltalk, its initial focus.
Several extensions covered other OOP languages with varying degrees of success. 
For example, \ct{Enum} and \ct{Class} were both \ct{Types} (see Figure \ref{fig:tyranny}, right).
In Java, both accept an \ct{implements} relationship to \ct{Interface} (another type).
Therefore this relationship is defined at the level of the \ct{Type} concept.
But then, a \ct{PrimitiveType} (also a \ct{Type}) inherits this same property which does not make sense in Java, and also an \ct{Interface} is allowed to \ct{implements}  another  \ct{Interface}, which again is not correct in Java.

Other generic meta-models have the same issue.
In the Dagstuhl Middle Meta-model \cite{Leth04a} (see Figure~\ref{fig:tyranny}, left), all relationships (\eg \ct{IsMethodOf}, \ct{Includes}, \ct{Invokes}) may occur between \ct{ModelElements} (\eg \ct{Type}, \ct{Routine}, \ct{Field}).
As a result many entities (\eg \ct{Field}) accept relationships (\eg \ct{IsMethodOf}) that they don't use.

A generic single inheritance meta-model thus turns out to be too permissive with entities owning properties that they don't use.
Incidentally, this has consequences on the size of the models since these properties occupy uselessly memory that multiplied by the number of instances (tens or hundreds of thousands) adds up to significant memory loss. 

EMF (Eclipse Modeling Framework) attempts to solve the problems raised by single inheritance inheritance tree. EMF implementation uses inheritance and interfaces, thus mimicking multiple inheritance.
But this still imposes to choose one dominant decomposition for inheritance and the other decompositions have to implement interfaces' API.

\subsection{ModMoose's Goals}

To design a new generation and modular reverse engineering environment we set goals at two different levels: meta-modeling and tooling.

\paragraph{Meta-modeling goals.} We are looking for generic modeling that leaves open a wealth of analyses (metrics, dead-code, design patterns, anti-patterns, dependency analysis, etc).
So we needed a unified representation that:
\begin{enumerate}
\item Supports the precise representation of various programming language features (identified in \cite{Shat19a});
\item Is compatible with many analysis tools (the \emph{processing environment} identified in \cite{Wash16a});
\item Is wary of memory consumption (identified in \cite{Lava10b});
\item Is extensible (allow for easy addition of new languages).
\end{enumerate}

\paragraph{Tooling goals.} Tools are needed to \cite{Bell98a,Kien10a}:
\begin{itemize}
\item Automate tedious tasks to bring tremendous speed-up with less errors.\footnote{A source code model computes in seconds a method call graph that takes weeks to recover by hand. We had the case in two different companies.}

\item Handle large quantity of data and in our case all the details of multi-millions lines of code.
If information size can become an issue even for automated tools, this is several orders of magnitude above what a human can handle.

\item Summarize data to allow a human abstracting a big picture understanding.

\item Quickly verify hypotheses so that there is little or no cost attached to wrong hypotheses, and several concurrent hypotheses can easily be compared.
\end{itemize}

\section{A Composable Meta-model of Programming Languages}
\label{sec:famixNG}

\famixng is a redesign of Famix around the idea of composing entities from language properties represented as traits \cite{Teso20b}.
With \famixng (See Section~\ref{famixng}), one defines a new meta-model out of reusable elements describing elementary concepts (See Section~\ref{creating}). 

\subsection{\famixng}\label{famixng}
To support the reuse of elementary language concept, \famixng relies on \emph{stateful traits}~\cite{Teso20b}.

\paragraph{What is a trait?} ``A trait is a set of methods, divorced from any class hierarchy. Traits can be composed in arbitrary order. The composite entity (class or trait) has complete control over the composition and can resolve conflicts explicitly''.
In their original form, traits were stateless, \ie pure groups of methods without any attributes.
Stateful traits extend stateless traits by introducing a single attribute access operator to give clients of the trait control over the visibility of attributes \cite{Teso20b}.
A class or another trait is composed from traits and it retains full control of the composition being able to ignore or rename some methods of the traits it uses.

In \famixng, all properties that were previously defined in classes are now defined as independent traits:
For example,  the trait \ct{TNamedEntity} only defines a \ct{name} property (a string) and may belong to trait \ct{TNamespaceEntity}.
Therefore any entity using this trait will have a name and can be part of a namespace entity.
Similarly, entities (presumably typed variables or functions) composed with the trait \ct{TTypedEntity} have a \ct{declaredType} pointing to an entity using the trait \ct{TType}. \\

Four types of traits  can be used to compose a new meta-model:
\begin{description}
\item[Associations] represent usage of entities in the source code.
This includes the four former associations of Famix: inheritance, invocation (of a function or a method), access (to a variable), and reference (to a type).
In \famixng, we also found a need for three more specialized associations: \ct{DereferencedInvocation} (call of a pointer to a function in C), \ct{FileInclude} (also in C), and \ct{TraitUsage}.

\item[Technical] traits do not model programming language entities but are used to implement Moose functionality.
Currently, this includes several types of \ct{TSourceAnchors}, associated to \ct{TSourcedEntity}to allow recovering their source code (a typical \ct{TSourceAnchor} contains a filename, and start and end positions in this file.)
Other \emph{Technical traits} implement software engineering metric computation, or are used to implement the generic query engine of Moose (see Section \ref{sec:tools-infra}).
There are 16 \emph{Technical traits} currently in \famixng.

\item[Core] traits model composable properties that source code entities may possess.
This includes \ct{TNamedEntity} and \ct{TTypedEntity} (see above), or a number of entities modeling ownership: \ct{TWithGlobalVariables} (entities that can own \ct{TGlobalVariables}), \ct{TWithFunctions} (entities that can own \ct{TFunctions}),... 
There are 46 \emph{Core traits} currently in \famixng including 38 traits modeling ownership of various possible kind of entities.

\item[Terminal] traits model entities that can be found in the source code such as \ct{Functions}, \ct{Classes}, \ct{Exceptions}, \ldots
These entities are often defined as a composition of some of the \emph{Core traits}.
For example, \ct{TClass} is composed of: \ct{TInvocationsReceiver} (class can be receiver of static messages), \ct{TPackageable}, \ct{TType} (classes can be used to type other entities), \ct{TWithAttributes}, \ct{TWithComments}, \ct{TWithInheritances}, \ct{TWithMethods}.
The name \emph{Terminal trait} refers to the fact that they can be used directly to create a programming language concept (a class, a package), whereas \emph{Core traits} are typically composed with other traits to make a meaningful programming language concept.
There are 38 such \emph{Terminal traits} currently in \famixng.
\end{description}

Contrary to the old Famix, the \emph{Terminal traits} are rather minimal definitions (intersection of languages) than maximal ones (union of languages).
They are intended to have only the properties found in any programming language.
For example \ct{TClass} does not use the \ct{TWithModifiers} trait (for attaching visibility) since not all languages explicitly define visibility of classes.
For any given meta-model, additional properties may be added to these ``\emph{Terminal traits}'', for example \ct{JavaClass} uses the traits \ct{TClass} and \ct{TWithModifiers} since Java classes can be declared public, private, \ldots
More specialized languages (\eg SQL) can always compose new entities from the set of \emph{Core traits} offered by \famixng.

We validated this new approach by defining different meta-models (\eg  GUI meta-model \cite{Verh19aShort}, Java, OpenStack cloud infrastructure, Pharo, PowerBuilder, SQL/PLPGSQL).

\subsection{Creating a new Meta-model with \famixng}
\label{creating}

To create a new meta-model, the developer may extend an existing meta-model or start from scratch.
This later case can be relevant if the language to model is quite different from what \modmoose currently handles, such as a specific domain or SQL. For example, while being different from procedural languages, stored procedures in SQL bear resemblance to functions and can be composed from the same \famixng traits.

\begin{figure}
\begin{minipage}[b]{0.6\linewidth}
\begin{lstlisting}[basicstyle=\scriptsize\ttfamily]
  entity := builder newClassNamed: 'Entity'.
  file := builder newClassNamed: 'File'.
  commit := builder newClassNamed: 'Commit'.
  author := builder newClassNamed: 'Author'.
  
  "inheritance"
  file --|> entity.
  file --|> TNamedEntity.
  commit  --|> entity.
  author  --|> entity.
  author --|> TNamedEntity.

  "properties"
  commit property: 'revision' type: #Number.
  commit property: 'date' type: #Object.
  commit property: 'message' type: #String.
  
  "associations"
  file *-* commit.
  commit *- author.
\end{lstlisting}
\end{minipage}
\begin{minipage}[b]{0.4\linewidth}
\raisebox{-1ex}{\includegraphics[width=\linewidth]{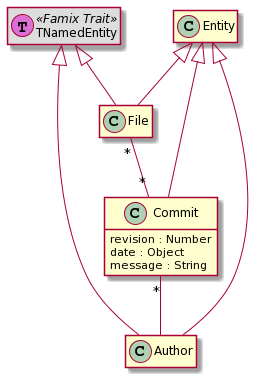}}
\end{minipage}
\caption{A simple File/Commit/Author meta-model (left: building script; right: UML view; gray: predefined \famixng trait)}
\label{fig:cvs-model}
\end{figure}

Figure~\ref{fig:cvs-model} shows the definition of a toy meta-model (left) with the meta-model itself (right).
There are four entities:  the root \ct{Entity}, and \ct{File}, \ct{Commit} and \ct{Author}.
In  grey, the \ct{TNamedEntity} trait, part of \famixng.
The embedded DSL needs first the entity creation, then definition of inheritance relationships, properties of each entity, and finally the relationships between entities.
To improve readability, entity descriptions are stored in variables having the same name (\eg\  \ct{author} variable for \ct{Author} entity description).
One can see an example of using a predefined trait, \ct{TNamedEntity} that adds a ``\ct{name : String}'' property to the classes using it.
The same syntax, ``\texttt{--|>}'', is used for class inheritance and traits usage.
From this, the builder generates all the corresponding classes and methods.

\paragraph{Meta-model reuse and extension.}
The extension of existing meta-model is done by specifying which meta-model to extend (similarly to \ct{import} instruction in Java), then the existing traits can be reused at will.
It is also possible to manage multi-language systems by composing several meta-models.
We cannot illustrate these two points for lack of space.

\section{\modmoose: A Reverse Engineering Environment}
\label{sec:moose-ide}

Program comprehension is still primarily a manual task.
It requires knowledge on computer science, the application domain, the  system, the organization using it, etc  \cite{Anqu07a}.
Yet the sheer size of the current software systems (millions of lines of code, hundred of thousands of entities modeled) precludes any software development team to fully understand these systems.
Any significant program comprehension activity imposes the use of specialized tools. 

\modmoose lets software engineers design specialized reverse engineering tools by (1) using infrastructure tools (see Section~\ref{sec:tools-infra}); (2) taking advantage of a bus architecture supporting smart interactions between tools (see Section~\ref{sec:tools-archi}); and, (3) reuse a set of generic and configurable tools (see Section~\ref{sec:tools-spec}).

\subsection{\modmoose Architecture}
\label{sec:moose-archi}

\begin{figure}[htbp]
  \begin{center}
  \includegraphics[scale=0.45,angle=-90]{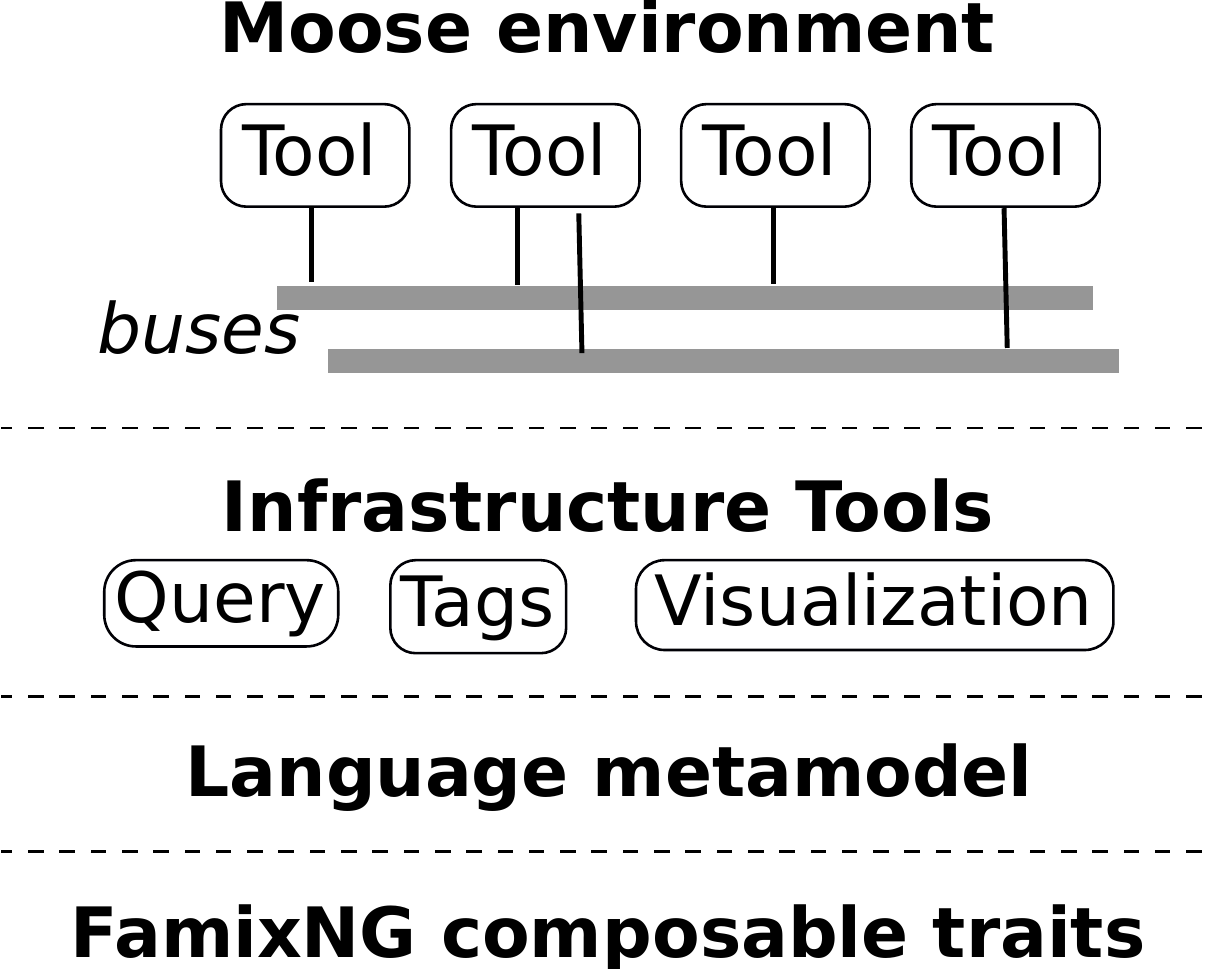}
  \caption{\modmoose architecture.}
  \label{fig:applicationArchitecture}
  \end{center}
\end{figure}

\modmoose is architectured on the following principles:
\begin{itemize}
  \item Tools are part of the \modmoose environment which acts as a master and centralizes data;
  \item Tools communicate through buses, they ``read'' model entities on their bus(es) and ``write'' entities back on their bus(es) (see below);
  \item Tools are focusing on a single task: \eg the Query Browser works on a set of model entities and produces another set of entities.
  \end{itemize}

These simple principles ensure that tools can be easily added to \modmoose and collaborate between themselves in a flexible manner.

\subsection{Infrastructure Tools}
\label{sec:tools-infra}

We identified three important requirement to analyze software systems:
(i) query and navigate a model to find entities of interest;
(ii) visualise the software to abstract information; and
(iii) annotate entities to represent meaning and reach a higher level of abstraction.

\subsubsection{Moose Query} is an API to programmatically navigate \famixng models.
For any \famixng meta-model, Moose Query\footnote{\url{https://moosequery.ferlicot.fr/}} computes the answer to generic queries:
\begin{itemize}
\item containment: parent or children entities of a given type from a current entity.

\item dependencies: following incoming or outgoing associations from or to a given entity.
Dependencies are also deduced from parent/children relationships, \eg dependencies between methods can be uplifted to dependencies between their parent classes or parent packages.

\item description: all properties of a given entity.
\end{itemize}

In a \famixng meta-model, relationships denoting a containment are declared as such in the association part of the meta-model construction (see Figure \ref{fig:cvs-model}), \eg \texttt{method *-<> class} or \texttt{class <>-* method}.
Containment queries may go up and down the containment tree based on the expected entity type: \eg from a \ct{Method} one can ask to go up to any entity(ies) of type \ct{Class} owning it.

Navigating dependencies is based on  containment and association relationships.
It is possible to navigate through a specific association (\eg ``all invoked methods'') or all types of association (\eg ``all dependencies'') to other elements. Direction of navigation (incoming, outgoing) must be specified. Dependencies between children of entities can be automatically abstracted at the level of their parents.

\paragraph{Package level communication example.}
A frequent query is to find how packages interact with each other via method calls.
This can be done by iterating over all the methods of all the classes of a package, collecting what other methods they invoke.
From this one finds the parent classes and parent packages of the invoked methods.

With Moose Query, such a query is simply expressed:\\
\texttt{(aPackage queryOutgoing: FamixInvocation) atScope: FamixPackage} (\ie find all invocations stemming from \texttt{aPackage}, and raise the result at the level of the receiving package).

\paragraph{Stored procedures referencing a column example.}
In SQL, one may want to know all the stored procedures\footnote{functions directly defined inside the database management system} accessing a given column of a database table.
These stored procedures can directly reference columns, in the case of triggers, or contain SQL queries that reference the columns.
SQL queries contain clauses that themselves can contain other queries.
So, the answer can be computed from a given column of a table, by collecting the references targeting this column, and recursively analyze them to identify the ones that are inside a stored procedure.
Due to the possible nesting of SQL queries, this has to be a recursive process. 

With Moose Query, this query is simply expressed as follows: \\
\texttt{aColumn queryIncomingDependencies atScope: FamixSQLStoredProcedure} (\ie find all incoming dependencies to \texttt{aColumn} and raise the result at the level of the stored procedures. Incoming dependencies not stemming from a stored procedures are simply dropped here).

\paragraph{Analysis.}
These examples show that MooseQuery API is independent from the meta-model used.
Obviously, each query depends on the meta-model of the model it is applied (one cannot ask for methods in a SQL model).
In the first example (Java), there are different kind of dependencies in the model (invocation, inheritance, access, reference), therefore the kind one is interested in must be specified (\texttt{queryOutgoing: FamixInvocation}).
In the second example there exist only one kind of dependencies to a column.
Consequently, the query can be simpler: \texttt{queryIncomingDependencies}.

In the second example, Moose Query implicitly filters out all queries that do not stem from a stored procedure. In the first example, no such filtering occurs since all methods belong to a package in Java.

\subsubsection{Visualisation Engines:}

\modmoose uses Roassal \cite{Berg16c}, a generic visualization engine, to script interactive graphs or charts.
Roassal is primarily used to display software entities in different forms or colors and their relationships.
Possible examples are to display the classes of a package as a UML class diagram, or as a Dependency Structure Matrix\footnote{\url{https://en.wikipedia.org/wiki/Design_structure_matrix}}.
Roassal visualization are interactive.

\modmoose also uses the Telescope, more abstract, visualization engine that eases building new visualizations in terms of entities, their relationships, their forms and colors, positioning, etc \cite{Larc15a}.
Telescope offers abstractions such as predefined visualizations and interaction, and relies on Roassal to do the actual drawing.
For example, Telescope offers a predefined ``Butterfly'' visualization, centered on an entity and showing to the left all its incoming dependencies and to the right all its outgoing dependencies.

\subsubsection{Tags}
are labels attached by the user to any entities either interactively or as result of queries \cite{Govi17a}.
Tags enrich models with extra information.
They have many different uses:
\begin{itemize}
\item to decorate entities with a virtual property \eg tagging all methods directly accessing a database.
\item to represent entities that are not directly expressed in the programming language constructs: \eg representing architectural elements or tagging subsets of a god class' members as belonging to different virtual classes \cite{Anq19aShort}.
\item to mark \textit{to do} and \textit{done} entities in a repetitive process.
\end{itemize} 

An important property of \modmoose tags is that they are first class entities.
Tags are not only an easy way to search entities, they can also be manipulated as any other model entity: a query may collect all dependencies from or to a tag, the visualization engines may display tags as container entities with actual entities within them, one can compute software engineering metrics (\eg cohesion/coupling) on tags considered as virtual packages.

\subsection{Smart Tool Interactions through Buses}
\label{sec:tools-archi}

The power of \modmoose comes from the generic interaction of specialized tools.
In particular, it is key that: (i) different tools display the \emph{same entities} from different points of view, or (ii) different instances of the same tool may be used to compare  different entities from the same point of view. 
\modmoose supports such scenarios using an open architecture based on buses and tool behavior controls.

\paragraph{Communication Buses.}
Tools communicate through buses.
They read entities produced by other tools on the buses and write entities on the same buses for other tools to consume.
Several buses can be created, and groups of tools can be set on different buses.
This allows one to explore different parts of the model or different courses of actions: \eg imagine two buses each one attached to a Query browser and a Dependency Graph browser (described in Section \ref{sec:tools-spec}).
Each Query browser selects different entities from the model, the Dependency Graph browsers display them, and the two buses ensure interaction within each pair of browsers.

Tools can also be detached from all available buses to keep their current result and presumably allow one to compare it with other results from other instances of the same tools.

Since tools can be attached to more than one bus, a tool can be set as a bridge between all buses: It listens to all buses and forwards activity on them.
A natural candidate for this is the Logger (described in Section \ref{sec:tools-spec}).
By selecting an interesting set of entities in the Logger, one can propagate these entities to all buses, thus synchronizing them.

\paragraph{Tool Behavior Controls.}
On top of the bus architecture, three behavior controls (\emph{frozen}, \emph{highlighting}, \emph{following}) fine-tune the tool reaction to bus events:

\begin{itemize}
\item \textit{Following} is the default state where a tool reads entities that pass on its bus(es) and writes entities to the same bus(es).
In such mode, a tool reacts to read entities. 

\item \textit{Frozen} is a state where the tool does not listen to the bus(es) for incoming entities and therefore keeps its state and display independent of entities written on the bus(es).
The tool is not detached from bus(es) and can still write entities on them.
A reverse engineer can interact with a frozen tool and other tools will be informed of this.

\item  \textit{Highlighting} is a sort of intermediary state, where the current entities on which the tool works remain the same (similar to \textit{Frozen}), but when new entities pass on the bus(es), the tool looks for them in its ``frozen'' display and highlight them if they are present (they are ignored otherwise). 
This is useful to highlight a new result in the context of a preceding result that was already displayed in the tool.
\end{itemize}

\subsection{Specialized Reusable Tools}
\label{sec:tools-spec}

On top of the modular meta-model (Section~\ref{sec:famixNG}) and infrastructure tools (Section~\ref{sec:tools-infra}), \modmoose offers some specialized tools that answer different recurring needs in software analysis.

\begin{figure}[htpb]
	\begin{center}
	\includegraphics[width=0.80\linewidth]{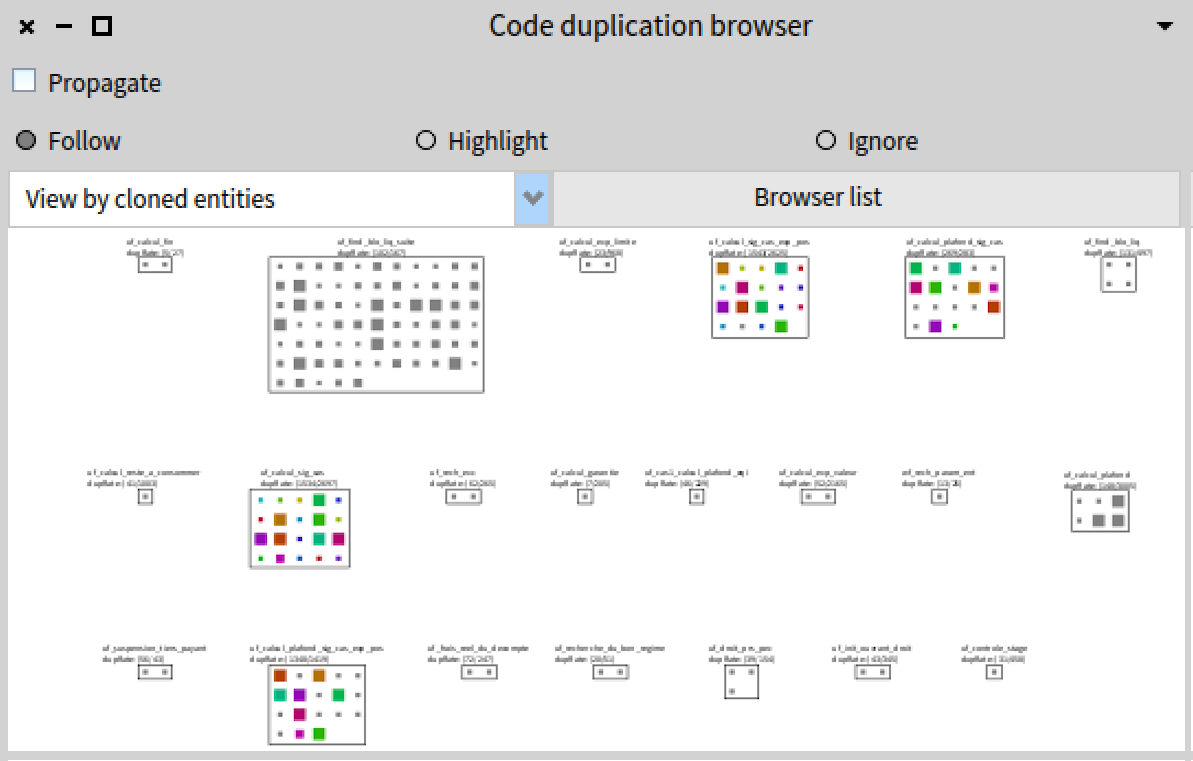}
	\caption{\modmoose  example of specialized tools: Duplication browser.}
	\label{fig:toolExamplesdup}  
	\end{center}
\end{figure}

\begin{figure}[htpb]
	\begin{center}
	\includegraphics[width=0.80\linewidth]{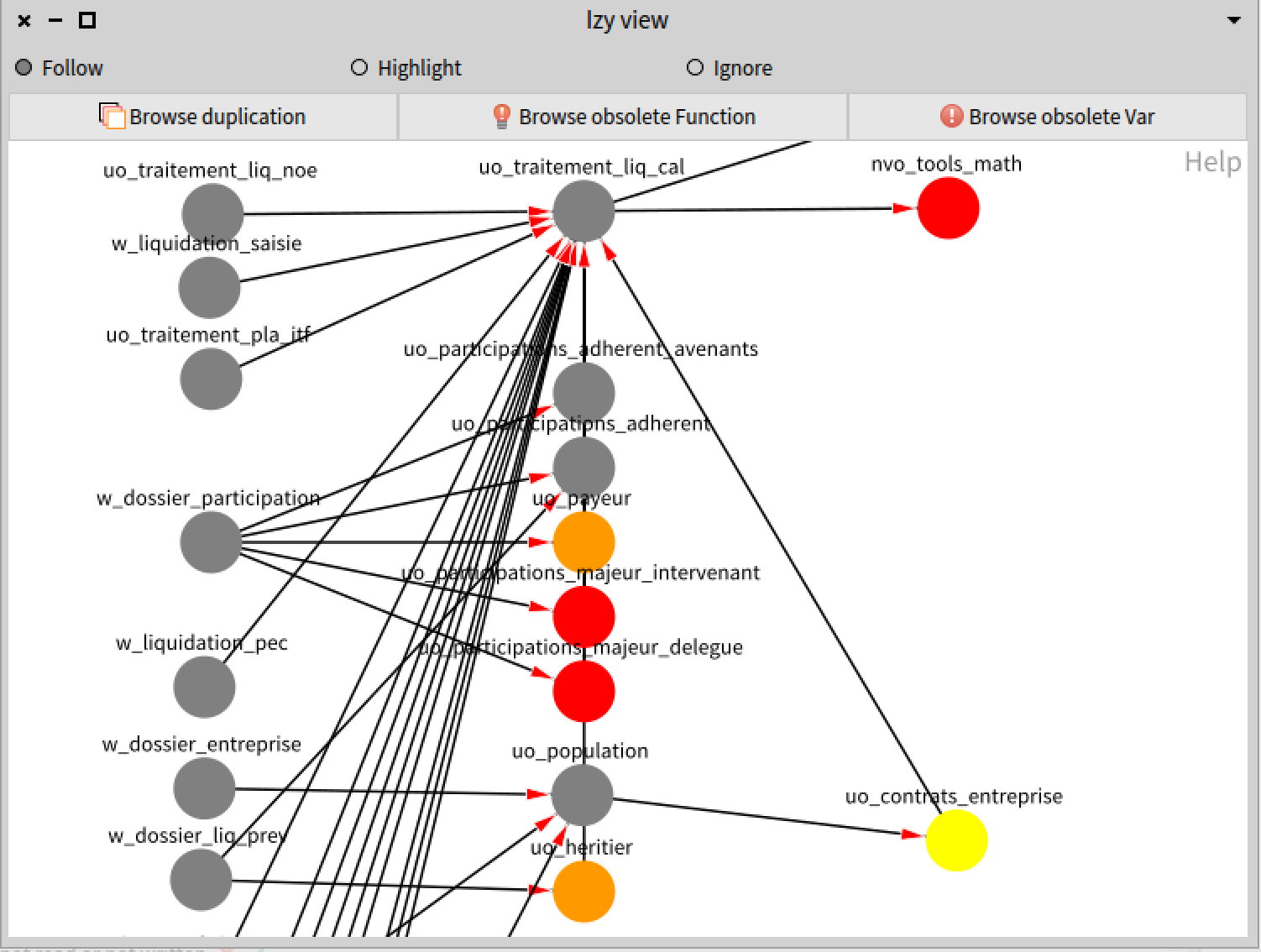}
	\caption{\modmoose  example of specialized tools: Call graph browser.}
	\label{fig:toolExamplescall}  
	\end{center}
\end{figure}

There is a number of reusable tools already implemented, among which:
\begin{description}
\item[Model Browser:] Imports or creates new models of given software systems, selects a model, and browses the model entities.
It outputs the selected model.
This is the entry point for working with the environment.

\item[Entity Inspector:] Lists all properties of a selected entity and their values (such as metrics, tags and others).
If the input is a group of entities, it lists all their common properties.
It allows navigating the model when the value of a property is another entity (\eg the \ct{methods} property of a \ct{Class}).
It outputs any entity selected for navigation.

\item[Query Browser:] Offers a graphical interface to the Query module (Section \ref{sec:tools-infra}).
It works on the entities found on the bus(es) and can filter and/or navigate from these entities.
It outputs the result of the queries.

\item[Dependency Graph Browser:] Shows as a graph all direct incoming and outgoing dependencies of a group of entities (see Figure~\ref{fig:toolExamplescall}, right).
Entities are selectable and are written on the bus for other tools to read them allowing interactive navigation.

\item[Duplication Browser:] Computes and displays duplication in the source code of a group of entities.
In Figure~\ref{fig:toolExamplesdup}, left, the big boxes are software entities (typically methods), and the small squares are the duplication fragments found in them.
Each duplication fragment has a color to identify it in all its occurrences (in the various software entities).
Entities are selectable and can be output on the bus(es).

\item[Source Code:] Displays the source code of an input entity.
If there are several entities in input or an entity with no source code (a model for example), does not do anything.
Currently produces no output.

\item[Logger:] Records all entities (individually or in a group) that pass on a bus.
It lets the user come back to a previous stage by selecting the entities that were produced at that stage.
It can export entities in files (txt, csv).
It outputs any selected entity or group of entities.
\end{description}

\subsection{Creating a new Tool}
New tools can be added and are planned, for example, a metric dashboard with standard metrics on the set of entities given in input, or new software maps such a Distribution Map visualisation \cite{Duca06c}.

Since tools are specialized, they are easy to develop, particularly with the help of the visualization engines and the query module. The open architecture with the buses makes the integration of new tools smooth and easy. 

\section{Conclusion}
\label{sec:conclu}

From literature and our industrial experience, we identified key aspects that a program comprehension environment must fulfill: a meta-model easily extensible and adaptable to represent new programming languages or sources of information; interoperating tools that can be adapted to the comprehension task at hand.

In this paper, we presented the new architecture of \modmoose, our reverse engineering environment.
It is first based on \famixng a new way to express meta-models by composing new entities from a set of traits, each describing individual properties that are generally encountered in programming languages.
We developed \emph{infrastructure} tools to manipulate models and an architecture with specialized end-user tools that interact through information buses.

This is not the end of the road.
We will complete our tool suite to respond to other aspects of reverse engineering and re-engineering.
Two research directions are drawing our attention:
(1) how to more easily develop importers for new programming languages;
(2) how to generate new code from the models in any programming language for which we have a meta-model.
\bibliographystyle{abbrv}
{\small\bibliography{rmod,others,local,short}}

\end{document}